\documentclass[%
 reprint,
showpacs,
 amsmath,amssymb,
 aps,
]{revtex4-1}

\usepackage{graphicx}
\usepackage{dcolumn}
\usepackage{bm}

\begin{document}

\title{Ultrafast Optical Modulation by Virtual Interband Transitions.}
\author{Evgenii E. Narimanov}
\affiliation{School of Electrical and Computer Engineering and Birck Nanotechnology Center, Purdue University, West Lafayette, Indiana 47907, USA}
\date{\today}

\begin{abstract}
A new frontier in optics research has been opened by the recent developments in non-perturbative optical modulation in both time and space that creates temporal boundaries generating ``time-reflection'' and ``time-refraction''  of light in the medium.  The resulting formation of a Photonic Time Crystal within the modulated 
optical material  leads to a broad range new  phenomena with a potential for practical applications, from  non-resonant light amplification and tunable lasing, to the new regime of quantum light-matter interactions. However,  the formation of the temporal boundary for light relies on optical modulation of the refractive index that is both strong and fast even on the time scale of a single optical cycle. Both of these two problems are extremely challenging even when addressed independently, leading to conflicting requirements for all existing methods of optical modulation. However, as we show in the present work, an alternative approach based on virtual interband transition excitation, solves this seemingly insurmountable problem. Being fundamentally dissipation-free, optical modulation by virtual excitation does not face the problem of heat accumulation and dissipation in the material, while the transient nature of the excited virtual population that modifies the material response only on the time scale of a single optical cycle, ensures that the resulting change in the refractive index is inherently ultrafast. Here we develop the theoretical description of the proposed modulation approach, and demonstrate that it can be readily implemented using already existing optical materials and technology.

\end{abstract}

\maketitle

\section{Introduction}
A large change in the refractive index of an optical material generally relies on introducing a substantial change in the energies of its electrons, whether bound or free.\cite{BornWolf,Boyd,Khurgin-index}  While this can be achieved in many different ways, from mechanical strain \cite{strain} and acousto-optics \cite{acousto-optics} to  thermal effects \cite{thermal} to  carrier density modulation, \cite{electrooptics} it is generally followed by the relatively slow dissipation (or recycling) of the resulting increase of the energy density (on the order of $10^4$ J/cm$^3$).\cite{Khurgin-index} The corresponding relaxation times (such as e.g. heat diffusion time, electron-hole recombination time, hot carrier recombination time, etc.) generally exceed sub-picosecond time scale, \cite{Khurgin-index} and do not enter the few femtoseconds range necessary for the formation of an optical Photonic Time Crystal.\cite{PTC1,PTC2,VSMotiR1,VSMotiR2} This situation is however dramatically reversed when free carriers undergo virtual interband transitions, induced by a strong optical pump pulse at the frequency just below the interband absorption range of the material, with only a few optical cycles in its duration $T$. While having insufficient energy to allow for a real transition of a free carrier  to another band, the inherent energy uncertainty in the pump photons, 
\begin{eqnarray}
\delta\varepsilon \simeq \frac{\hbar}{T}
\end{eqnarray}
may provide the required ``excess energy'' to promote the interband transition and ensure the formation of virtual populations, that necessarily modify the refractive index of the material. Note that this is an inherently transient response, as such virtual populations are only manifest in the presence of the pump pulse -- leading to nearly-instantaneous ``turn-on'' and ``turn-off'' of the refractive index modulation.

As  virtual transitions are generally  dissipation-free,\cite{LLQM} the proposed approach is free from the energy dissipation/recycling bottleneck, that limits the 
electromagnetic energy density used to change the dielectric permittivity of the material and thus magnitude of 
refractive index variation in the conventional methods.\cite{Khurgin-index}  Without the need to allow for the slow relaxation of the refractive index, it can be periodically modulated with the period that is comparable to that of a single optical pump pulse duration -- which therefore offers a clear and immediate solution of the problem of 
ultrahigh frequency periodic modulation of the refractive index, that in necessary for the band formation in the elusive optical 
Photonic Time Crystal.\cite{VSMotiR1}

\section{The Concept}

\begin{figure}[htbp] 
 \centering
 \includegraphics[width=3.5in]{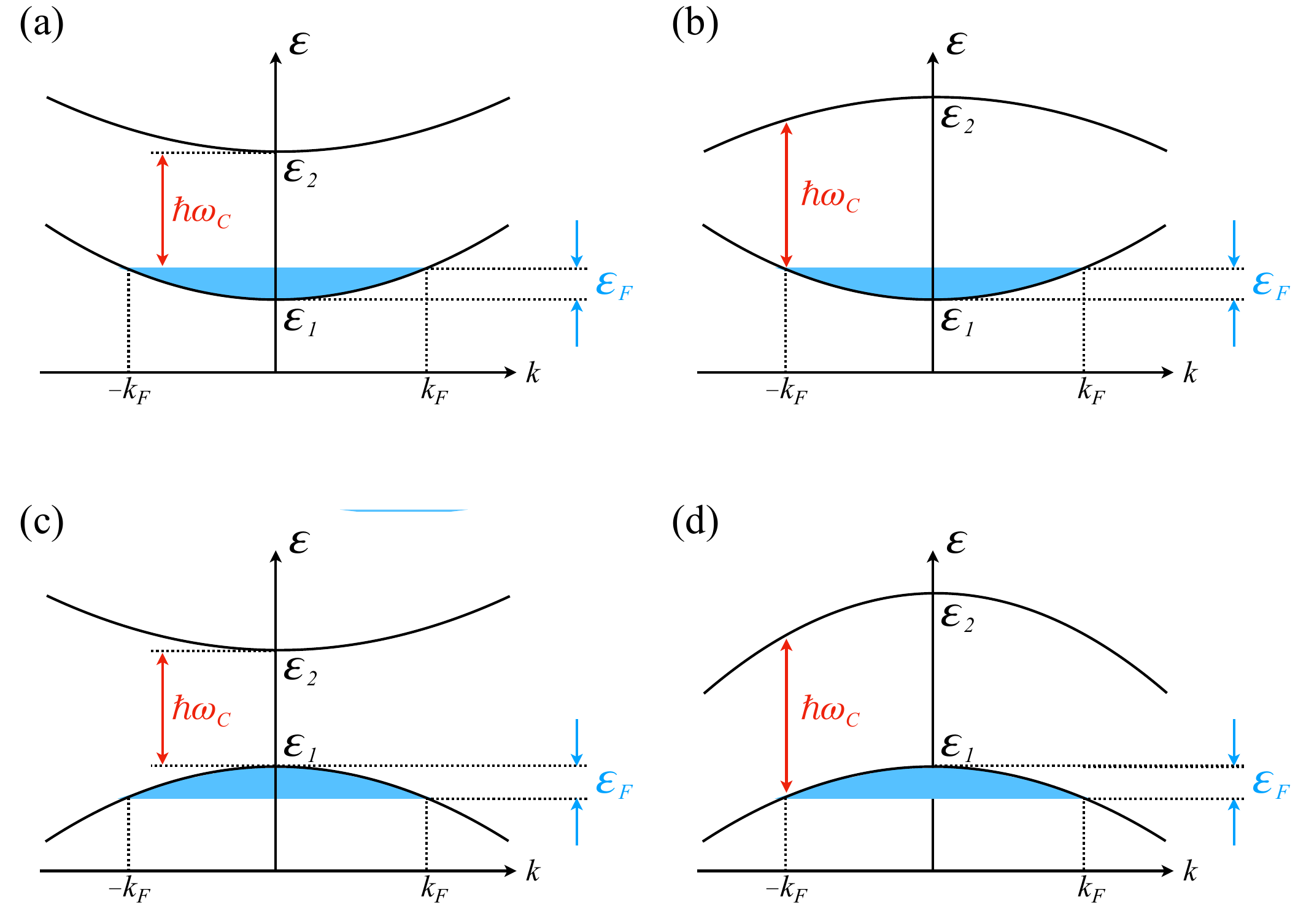} 
   \caption{  The schematic representation of the bandstructure  of the conducting optical material. for different combinations of the free carrier effective masses in the 
   two bands: (a) $m_1^* > 0$ and $m_2^* > 0$,  (b) $m_1^* > 0$ and $m_2^* <  0$, (c) $m_1^* < 0$ and $m_2^* > 0$, (d) $m_1^* < 0$ and
   $m_2^* <  0$. The blue 
   color filling represents the 
   equilibrium free carriers populations -- electrons for (a) and (b), and holes for (c) and (d). $\varepsilon_1$ and $\varepsilon_2$ are the band edge energies, 
   $\varepsilon_F$ and $k_F$ are the Fermi energy and Fermi wavenumber of the free carriers, $ \omega_c$ is the interband absorption edge frequency.   }
   \label{fig:bands}
\end{figure}

The proposed approach to ultrafast optical modulation only relies on the presence of a sufficient number of free carriers in either the conduction or valence band of the optical material, and it equally applicable to all band topologies (see Fig. \ref{fig:bands}). While it may be advantageous to choose the operating frequency $\omega_S$ above the free carrier  plasma frequency $\Omega_p$ (to ensure relative transparency of the material) or close to $\Omega_p$ (for the operation in the epsilon-near-zero regime when a given absolute variation in the refractive index turns into a higher relative change), these  are neither  essential nor necessarily advantageous for the implementation of the present concept of ultrafast refractive index modulation.

In the proposed approach, the material is optically pumped with the ultra-short pulse (of duration $T$ that is on the order of few of its optical cycles) at the frequency $\omega_P$ below the interband absorption edge $\omega_c$ (see Fig. \ref{fig:bands}), with
\begin{eqnarray}
\hbar \left( \omega_c - \omega_P \right) \simeq \frac{\hbar}{T}.
\label{eqn:regime}
\end{eqnarray}
In this regime, the pump photon energy uncertainty $\Delta\varepsilon \sim \frac{\hbar}{T}$
 is comparable to the energy difference between the band offset and the pump photon energy $\hbar\omega_P$, leading to a substantial virtual 
 carrier population formation in different bands. This excited carrier population is inherently transient, and can only exist for no longer that few optical cycles, while the pump pulse is present. The pump-induced virtual populations are therefore only manifest for the duration of the ultra-short pump pulse, regardless of the mechanism of the band-to-band relaxation/recombination in the materials and its corresponding time scale $\tau_1$. This inherently ultrafast and dissipation-free nature of the virtual interband transitions -- the capability that's essential for formation of the photonic time crystals \cite{VSMotiR1}  -- is an  essential feature of the proposed refractive index modulation approach. 

The proposed concept of the strong modulation of the dielectric permittivity due to the formation of transient  populations, 
relies on the material response where virtual carriers contribute to the electromagnetic response of the medium in essentially the same manner as real electrons (albeit without relaxation).
While possibly counter-intuitive, this behavior  has been firmly established for nearly half a century, since it was discovered in mid-1980s in the ultrashort pulse generation experiments with GaAs quantum wells, where 
intense optical interaction below the optical absorption threshold induced substantial virtual populations with measurable consequences.\cite{Yamanishi-PRL-1987},
In particular, measurements of ac Stark effect in GaAs quantum wells showed clear evidence of  unharmonic interactions of virtually created electrons and holes.\cite{Mysyrowicz-PRL-1986},\cite{VLCZH-OL-1986}
 Furthermore, the resulting screening of the  virtual carrier populations was essential for 
the generation of ultrashort electrical pulses in biased quantum wells.\cite{CMSR-PRL-1987}
This well established experimental evidence firmly justifies the proposed mechanism where the virtual carriers respond to the electromagnetic field as if they are real electrons and holes,  leading to the strong and fast modulation of  the dielectric permittivity.

Furthermore, it has also been demonstrated that while virtual carrier populations produce exactly the same physical effect as the real ones,  they do not participate in any relaxation process \cite{SRC-PRL-1986} -- which is the key for our proposed ultrafast modulation scheme.

\section{The Model}

\begin{figure}[htbp] 
 \centering
 \includegraphics[width=2.5in]{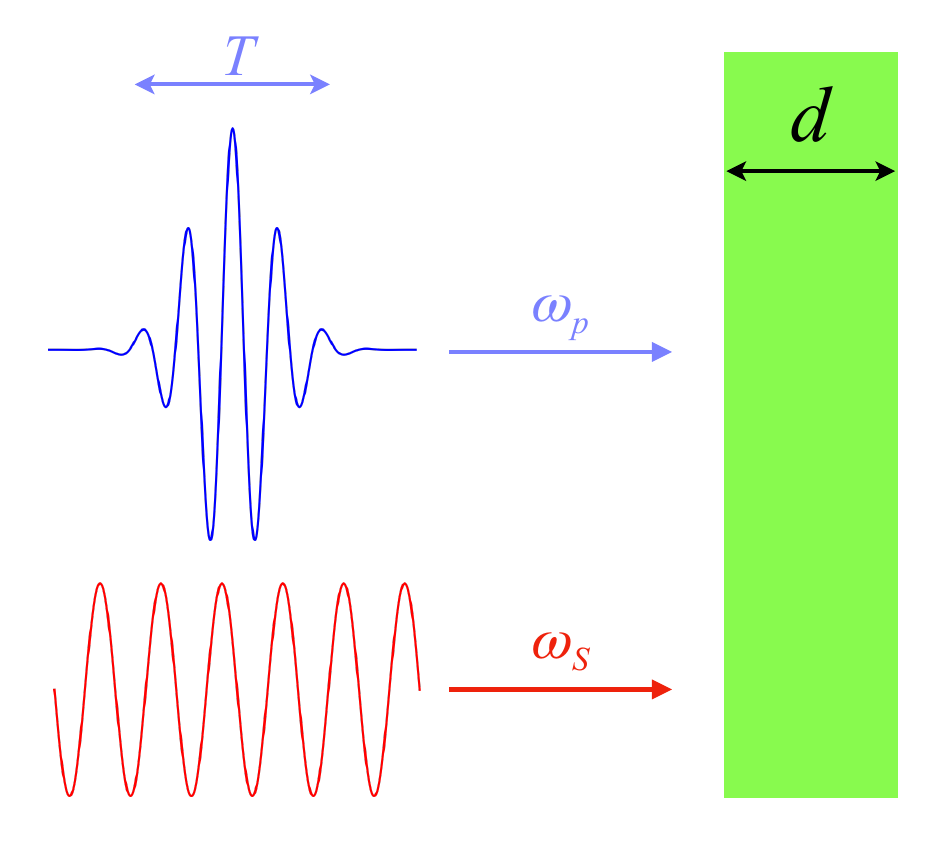} 
   \caption{The schematic representation of the proposed modulation approach. An optically thin layer of the conducting optical
   material (with thickness $d$) is illuminated by the signal (``probe'')  beam at the frequency $\omega_S$, and a high-intensity 
   ``pump'' pulse with the center frequency $\omega_P$ with the duration $T$ of only a few optical cycles.}
   \label{fig:schematics}
\end{figure}

We consider a parallel slab of an optical material that is simultaneously illuminated with with a high-intensity ``pump''   pulse at the central frequency $\omega_P$ with only a few optical cycles in its duration $T$, while the signal at the frequency $\omega_S$ operates in the linear response regime -- see Fig. \ref{fig:schematics}.
The material supports a sizable free-carrier population, characterized by the corresponding plasma frequency  $\Omega_P$ that
is comparable to $\omega_S$ (see Fig. \ref{fig:schematics}) but is well below the pump frequency $\omega_P$, that is in turn below interband absorption edge $\omega_c$ (indicated in Fig. \ref{fig:bands} for various band topologies):
\begin{eqnarray}
\Omega_p \lesssim \omega_S \ll \omega_P \lesssim \omega_c - \frac{1}{T}.
\label{eq:regime}
\end{eqnarray}
With the same order of magnitude for the operating frequency $\omega_S$ and the free carrier plasma frequency $\Omega_p$, a noticeable material absorption is unavoidable, which precludes the use of optically thick materials. We therefore assume the regime of subwavelength thickness of the optical material,\
\begin{eqnarray}
\frac{\omega}{c} d \ll 1,
\label{eqn:wcd}
\end{eqnarray}
which allows to neglect the retardation within the modulated material for both the ``pump'' $\omega_P$ and the ``probe'' $\omega_S$. 

\section{The Theory}

For the free carriers in the optical material subject to both pump and probe/signal fields, the corresponding von Neumann density function equations \cite{Boyd} take the form 
\begin{eqnarray}
i \hbar \frac{\partial \rho_{n_1 {\bf k}_1, n_2 {\bf k_2}}}{\partial t} & = &\left(  \varepsilon_{n_1{\bf k}_1}  -  \varepsilon_{n_2{\bf k}_2} \right) \rho_{n_1 {\bf k}_1, n_2 {\bf k_2}}\nonumber \\
& - & i \hbar \ \gamma_{n_1 {\bf k}_1, n_2 {\bf k_2}} \left(\rho_{n_1 {\bf k}_1, n_2 {\bf k_2}}- \rho^{(eq)}_{n_1 {\bf k}_1, n_2 {\bf k_2}} \right) \nonumber \\
& - & \left[ \hat{H}_E^{(P)} +  \hat{H}_E^{(S)}, \hat{\rho} \right]_{n_1 {\bf k}_1, n_2 {\bf k_2}} , \ \ \ \ \ \label{eqn:rho}
\end{eqnarray}
where $\hat{\rho}$ is the free-electron density matrix operator with the equilibrium value $\hat{\rho}^{(eq)}$, $\gamma$ is the matrix of  the corresponding (phenomenological) relaxation coefficients,  $n$ and ${\bf k}$ are the free electron band index and Bloch momentum,  while 
$\hat{H}_E^{(P)}$ and $\hat{H}_E^{(S)}$ are  the effective Hamiltonians of electron interaction with the (time-dependent) ``pump'' and ``signal'' electromagnetic  fields.

Note that, even within the assumptions of Eqn. (\ref{eqn:wcd}), one may still be tempted to ``fully'' account for the electromagnetic field spatial variation in the direction of propagation $\sim \exp\left(i q z\right)$ within the optical material (where $q$ is the electromagnetic wavenumber), leading to the corresponding spatial dispersion corrections in the resulting theory.\cite{LLECM} However, from plasmonic metals ($\Omega_p$ is the UV range \cite{SCai}) to transparent conducting oxides ($\Omega_p$ in near-infrared \cite{TCOs}) to highly doped semiconductors ($\Omega_p$ in mid-IR \cite{nmat}), the free carrier mean free path does not exceed a small fraction of the operating optical wavelength. Therefore, on the scale of the free electron wavefunctions in the scattering environment, the acting field in the matrix element for electronic interband transitions should be treated as uniform, leading to the selection rule for the ``vertical'' 
interband transitions ${\bf k}_1 = {\bf k}_2$, for $n_1 \neq n_2$,
\begin{eqnarray}
\big\langle n1, {\bf k}_1 \left| H_E^{(P)} \right| n_2, {\bf k}_2 \big\rangle_{n_1 \neq n_2}  & = & - e E_0\left(t\right)\  \delta_{{\bf k}_1{\bf k}_2} \ x_{n_1n_2}. \ \ \ \  \label{eqn:vertical} 
\end{eqnarray}  
Here $H_E^{(P)}$ is the effective Hamiltonian of light-matter interactions for the (time-dependent) ``pump field''
\begin{eqnarray}
{\bf E}\left({\bf r}, t\right) & = &E_0\left(t\right) {\bf \hat{e}} \cos\left(- i \omega_P t + i \phi_0\right), 
\end{eqnarray} 
$\delta$ is the (Kronecker's) delta-function,  ${\bf \hat{e}}$ is the unit vector that represents the field polarization, and the interband matrix element
\begin{eqnarray}
 x_{n_1n_2}\left({\bf k}\right) & \equiv & \frac{1}{v_0} \int_{v_0} d{\bf r} \ u_{n_1{\bf k}}\left({\bf r}\right) \  {\bf \hat{e}\cdot{\bf r}}  \ u_{n_1{\bf k}}\left({\bf r}\right), \ \ \ 
\end{eqnarray}
where $n$ is the band index, $u_{n {\bf k}}\left({\bf r}\right)$ is the free carrier Bloch wavefunction, and $v_0$ is the volume of the material unit cell. 

On the other hand, for the corresponding intra-band ($n_1 = n_2$) matrix element, the free-carrier Bloch functions can be adequately approximated by their values at the band energy extremum,\cite{Ziman} leading to
\begin{eqnarray}
\big\langle n1, {\bf k}_1 \left| H_E^{(P)} \right| n_2, {\bf k}_2 \big\rangle_{n_1 = n_2}  & = & - e {E}_0  {\bf \hat{x}} \cdot  \frac{\partial}{\partial{\bf k}_2} \  \delta_{{\bf k}_1{\bf k}_2}.\ \ \ \  \label{eqn:diagonal} 
\end{eqnarray}  

As the amplitude of the  ``signal'' $E_S\left(t\right)$  is small to ensure the linear-response regime operation, and the corresponding frequency is well below the inter-band absorption edge ($\omega_S \ll \omega_c$ -- see Eqn. (\ref{eqn:regime})), the solution of (\ref{eqn:rho}) in terms of $E_S\left(t\right)$ can be obtained using the standard perturbative approach,\cite{Boyd} leading to the  effective dielectric permittivity
\begin{eqnarray}
\epsilon\left(\omega_S,t\right)  & = & 
\epsilon_\infty\left(\omega_S\right) - \sum_n \frac{4 \pi }{\gamma_{nn} \ m_n^*} \int_{-\infty}^t dt' \ \Delta N_n\left(t'\right) \nonumber \\
& \times &  \left[1 - \exp\left( - \gamma_{nn} \left(t - t'\right) \right) \right] \exp\left( - i \omega_S t'\right),
\label{eqn:eps_S}
\end{eqnarray}
where $m^*_n$ is the free carrier effective mass in the band $n$, 
$\gamma_{n_1n_2} \equiv \langle \gamma_{n_1n_2} \rangle_{k < k_F}$ is the free carrier  relaxation rate averaged over the volume 
of the Fermi surface of the conduction band, and  the transient population density $\Delta N_n\left( t\right)$ is defined in terms of 
the diagonal elements of density matrix $\bar{\rho}$ of the free carriers strongly coupled to the pump pulse
\begin{eqnarray}
\Delta N_n\left(t\right) & = & \frac{2}{\left(2 \pi\right)^3} \int d{\bf k} \left(\bar{\rho}_{nn}\left({\bf k},t\right) - \bar{\rho}^{(eq)}_{nn}\left({\bf k}\right) \right) 
\label{eqn:DN}
\end{eqnarray}
where with  ``vertical transition'' selection rule of Eqn. (\ref{eqn:vertical}) the reduced density matrix $\hat{\overline{\rho}}$ satisfies the (infinite) system of coupled nonlinear differential equations
\begin{eqnarray}
i \hbar \frac{\partial \bar{\rho}_{n_1  n_2 }\left({\bf k},t\right) }{\partial t} & = &\left(  \varepsilon_{n_1{\bf k}_1}  -  \varepsilon_{n_2{\bf k}_2} \right) \bar{\rho}_{n_1, n_2} \left({\bf k},t\right) \nonumber \\
& - & i \hbar \ \gamma_{n_1  n_2 } \left(\bar{\rho}_{n_1 n_2} \left({\bf k},t\right)- \bar{\rho}^{(eq)}_{n_1  n_2}\left({\bf k}\right) \right)  \nonumber \\
& + & e E_P\left(t\right)\ \left[\bar{x}, \bar{\rho}\right]_{n_1n_2},
\label{eqn:rhoP}
\end{eqnarray}
with the effective dipole matrix element $e \bar{x}_{n_1n_2} \equiv \langle  e x_{n_1n_2}\left({\bf k}\right)  \rangle_{k < k_F}$ again defined in terms of the reciprocal space average over the volume of the Fermi surface of the conduction band.

The requirement of strong coupling between the optical pump pulse and the free carriers in the material, 
\begin{eqnarray}
\frac{\hbar}{T} \sim \hbar \left(\omega_c - \omega_P\right)
\label{eqn:strong}
\end{eqnarray}
can be only satisfied for a single pair of electronic bands (which will be represented by the indices $n_1=1$ and $n_2=2$,  for the occupied and unoccupied bands respectively -- see Fig. \ref{fig:bands}), which allows to neglect all other interband transitions in Eqn. (\ref{eqn:rhoP}). The system (\ref{eqn:rhoP})  then effectively
separates  into a set of uncoupled pairs of equations for  $\bar{\rho}_{21}\left({\bf k}\right)$ and the ``population inversion'' $w\left({\bf k}\right)  \equiv  \bar{\rho}_{22}\left({\bf k}\right)-\bar{\rho}_{11}\left({\bf k}\right)$:
\begin{eqnarray}
i \hbar \frac{d\bar{\rho}_{21}}{d t} & = &  \left(\Delta_k+ \frac{i}{\tau_2}\right) \bar{\rho}_{21}\left(t\right)  +  e \bar{x}_{21} E_S\left(t\right) w\left(t\right), \label{eqn:rho21} \\
 \frac{d w}{d t} & = & - \frac{w\left(t\right) - w^{(eq)}}{\tau_1} + \frac{4e\bar{x}_{21}}{\hbar}   E_S\left(t\right) {\rm Im}\left[ \bar{\rho}_{21}\left(t\right)\right] , 
 \label{eqn:w} \ \ \ \ \ \ 
\end{eqnarray}
where $\varepsilon_1$ and $\varepsilon_2$ are the electronic band edge energies (see Fig. \ref{fig:schematics}),  the ($k$-dependent) detuning factor
\begin{eqnarray}
\Delta_k & = & \frac{1}{\hbar }\left( \varepsilon_2 - \varepsilon_1 \right) - \frac{\hbar k^2}{2 m_1^*} \left(1 - \frac{m^*_1}{m_2^*}\right) - \omega_S,
\end{eqnarray}
and we've used the standard notation \cite{Boyd} for the population relaxation time $\tau_1 \equiv 1/\gamma_{11} = 1/\gamma_{22}$ and  the dephasing time $\tau_2 \equiv 1/\gamma_{12} = 1/\gamma_{21}$.

\begin{figure}[htbp] 
 \centering
 \includegraphics[width=3.25in]{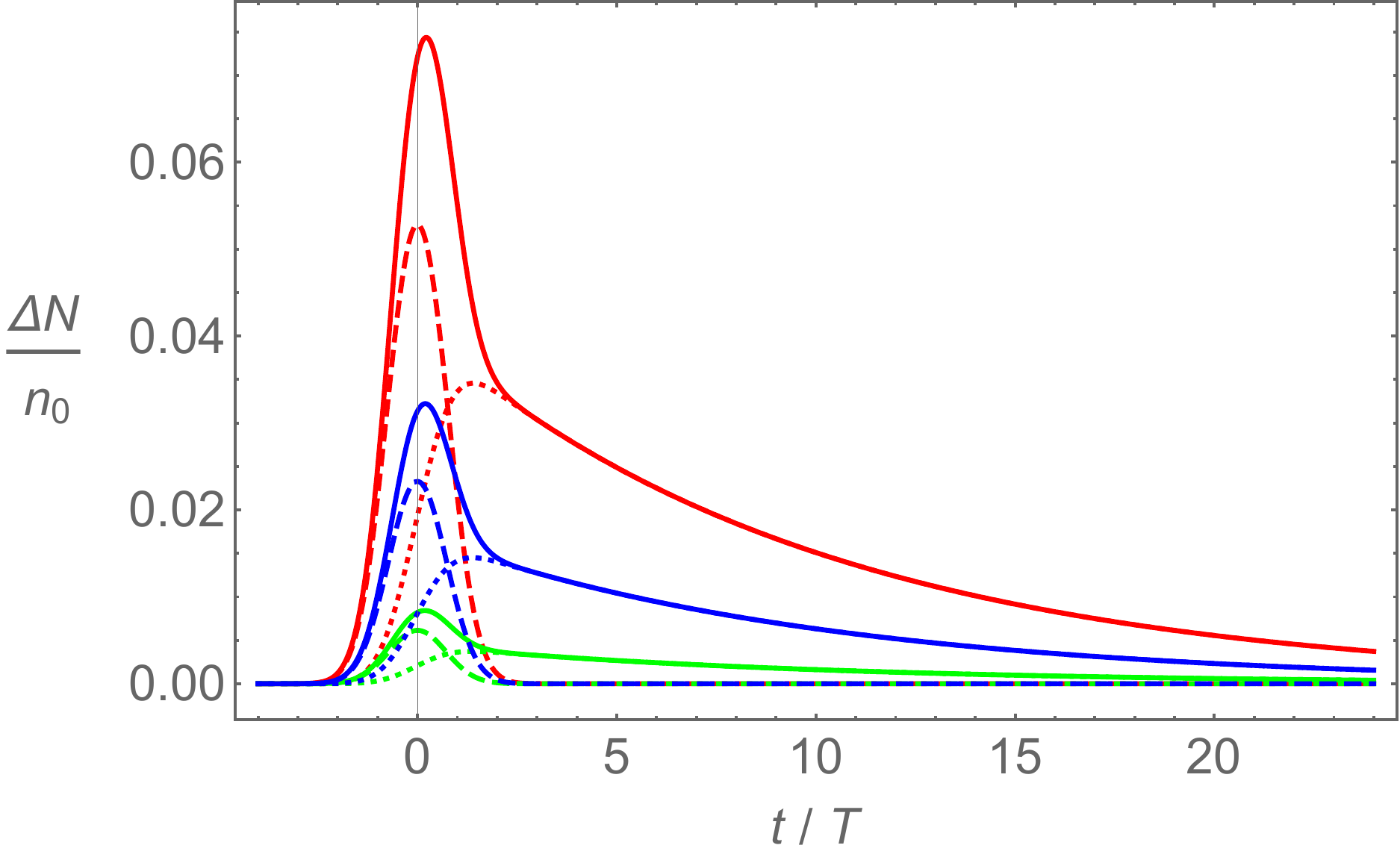} 
   \caption{The change in the population density $\Delta N\left(t\right)$, induced by the 
   optical pump pulse $I_P\left(t\right) = I_0 \exp\left( - t^2  / T^2 \right)$, for $\alpha_* = 0.25$ (red curves),
   $\alpha_* = 0.1$ (blue lines), and $\alpha_* = 0.025$ (green curves).
   Solid curves correspond to the total population change $\Delta N$, dashed lines are the virtual contributions
   $\Delta N_1$ and dotted curves correspond to  $\Delta N_2$. The relaxation times are $\tau_1 = 10 T$ and
   $\tau_2 = 5 T$.
   }
   \label{fig:populations}
\end{figure}

In the limit of (\ref{eqn:regime}) which is the focus of the present work, Eqns. (\ref{eqn:rho21}),(\ref{eqn:w}) can be solved analytically, which for thermal energy 
well below $\varepsilon_F$ and $\varepsilon_2 - \varepsilon_1$  yields
\begin{eqnarray}
w\left(t\right) & = & -\frac{1}{\sqrt{1+\alpha_k I_P\left(t\right)}} + \frac{1}{\Delta_k^2 \tau_1^2}  \exp\left(-\frac{t }{\tau_1}\right) \nonumber \\
& \times &\int_{-\infty}^t dt' \exp\left(\frac{ t'}{\tau_1}\right)
\frac{d}{dt'} \ \frac{1}{\sqrt{1+\alpha_k I_P\left(t'\right)}},
\end{eqnarray}
where $I_P\left(t\right) \equiv E_P^2$ is the optical pump intensity, and $\alpha_k  \equiv {\bar{x}^2_{21} }/{\Delta_k^2}$.
Substituting 
\begin{eqnarray}
\bar{\rho}_{11}& = & \left( 1 - \frac{w}{2} \right) \  \theta\left(k_F - k\right), \\
\bar{\rho}_{22}& = &   \frac{w}{2}  \  \theta\left(k_F - k\right),
\end{eqnarray}
into (\ref{eqn:DN}), for $\varepsilon_F < \varepsilon_2 - \varepsilon_1$ we obtain
\begin{eqnarray}
\Delta N\left(t\right) & = & \Delta N_1 + \Delta N_2,
\end{eqnarray}
where $\Delta N_1$ is the virtual induced population
\begin{eqnarray}
 \Delta N_1\left(t\right) & = & \frac{n_0}{2} \left( 1 - \frac{1}{\sqrt{1 + \alpha_* I_P\left((t\right)}} \right), 
\end{eqnarray}
and
\begin{eqnarray}
 \Delta N_2\left(t\right) & = &  \frac{n_0}{2} \int_0^{\infty} \frac{dt'}{\tau_2} \frac{\alpha_* I_P\left(t-t'\right)\ \exp\left( - t' / \tau_1\right)}{\sqrt{1 + \alpha_* I_P\left((t-t'\right)}}.
 \ \ \ \  
\end{eqnarray}
Here $n_0$ is the equilibirum free carrier density, and 
\begin{eqnarray}
\alpha_*  \equiv \frac{e^2 \bar{x}^2_{21} }{\left( \varepsilon_2 - \varepsilon_1 -
\frac{3}{5} \varepsilon_F \left(1 - \frac{m^*_1}{m_2^*}\right)
- \hbar\omega_S\right)^2}.
\end{eqnarray}
As expected, the virtual transitions lead to the formation of the transient population $\Delta N_1$ for the duration of the optical pump pulse, 
while the multi-photon absorption yields the ``real'' population change $\Delta N_2$ that persists for a much longer duration defined by
 the relaxation time $\tau_1$. This behavior is illustrated in Fig. \ref{fig:populations}.

For the dielectric permittivity at the signal frequency $\omega_S$ we then obtain
\begin{eqnarray}
\epsilon\left(\omega_S,t\right)  & = & \epsilon_\infty\left(\omega_S\right) 
- \frac{\Omega_p^{\rm (eff)}\left(t\right)^{\ 2}}{\omega_S \left(\omega_S + {i}{\gamma_{\rm eff}\left(t\right)}\right)} , 
\label{eqn:epsilon}
\end{eqnarray}
where $\epsilon_\infty\left(\omega\right)$ is the background permittivity due to the contributions of the bound electrons in the material,   the effective time-dependent plasma frequency
\begin{eqnarray}
\Omega_p^{\rm (eff)}\left(t\right)
& = &
\Omega_p \sqrt{1 -  \eta_1\left(t\right)  - \eta_2\left(t\right)},
\end{eqnarray}
where $\Omega_p$ is the original plasma frequency in the medium before the activation of the optical pump, and the effective loss rate
\begin{eqnarray}
\gamma_{\rm eff}\left(t\right) & = & \frac{ 1 + \nu\left(t\right)}{\tau_2}.
\end{eqnarray}
Here, the contribution $\eta_1\left(t\right)$  represents the relative change of the plasma frequency 
due to the formation of the virtual  carrier  population in the high-energy band at the expense of the carrier density in the conduction band for the duration of the optical pump
$I_P\left(t\right)$. As expected, the modulation of the 
effective plasma frequency $\Omega_p^{\rm (eff)}$ due to $\eta_1\left(t\right)$ is only present on the time scale $T$ of the optical pump pulse $I_S\left(t\right)$. In contrast to this behavior, ``real'' population change accounted for by $\eta_2\left(t\right)$ that is induced by the multi-photon absorption in the medium, persists for the duration of subsequent carrier relaxation time $\tau_1$.  We find
\begin{eqnarray}
\eta_1\left(t\right) & = &  \frac{1}{2} \left(1 - \frac{m^*_1}{m_2^*}\right)  \left\{ 1 - \frac{1}{\sqrt{1+\alpha_* I_P\left(t\right)}}  \right\}, \ \ \ 
\label{eqn:eta1}
\end{eqnarray}
and
\begin{eqnarray}
\eta_2\left(t\right) & = &  \frac{1}{2} \left(1 - \frac{m^*_1}{m_2^*}\right)  \left\{
 \int_{-\infty}^t \frac{dt'}{\tau_2} 
 \ \frac{\alpha_* I_P\left(t'\right)   \ e^{- \left( t - t'\right)/\tau_1}}{\sqrt{1+\alpha_* I_P\left(t'\right)}} \right. \nonumber \\
& + & \left.  \int_{-\infty}^t {dt'}  \  \exp\left(-\frac{  t - t'}{\tau_2} \right) \ \cos\left(  \omega_S\left( t - t'\right)\right) \right. \nonumber \\
& \times & \left. \frac{d}{dt'} \ \frac{1   }{\sqrt{1+\alpha_* I_P\left(t'\right)}}  \right\}, \
\label{eqn:eta2}
\end{eqnarray}
while
\begin{eqnarray}
\nu\left(t\right) & = &  
\frac{1}{2} \left(1 - \frac{m^*_1}{m_2^*}\right)   \int_{-\infty}^t 
\frac{dt'}{\tau_1}  \ \exp\left({- \frac{ t - t'}{\tau_1} } \right)
\nonumber \\
& \times & \left\{ \frac{\alpha_* I_P\left(t'\right)}{\sqrt{1+\alpha_* I_P\left(t'\right)}} 
+  \frac{\tau_1}{\tau_2}   \  \sin\left(\omega_S\left(t' - t\right)\right) \right. \nonumber \\
& \times & \left.  \left[  \frac{\omega_S \tau_2 \ \alpha_* I_P\left(t'\right)}{\sqrt{1+\alpha_* I_P\left(t'\right)}}
-
\frac{d}{dt'}  \frac{1}{\sqrt{1+\alpha_* I_P\left(t'\right)}} 
  \right]
\right\}. \ \ \ \ \ \ 
\label{eqn:nu}
\end{eqnarray}
 Note that the signs
of $\eta_{1,2}\left(t\right)$ and $\nu\left(t\right)$ depend on the difference between the free carrier effective masses in the bands. 

\begin{figure}[htbp] 
 \centering
\includegraphics[width=3.25in]{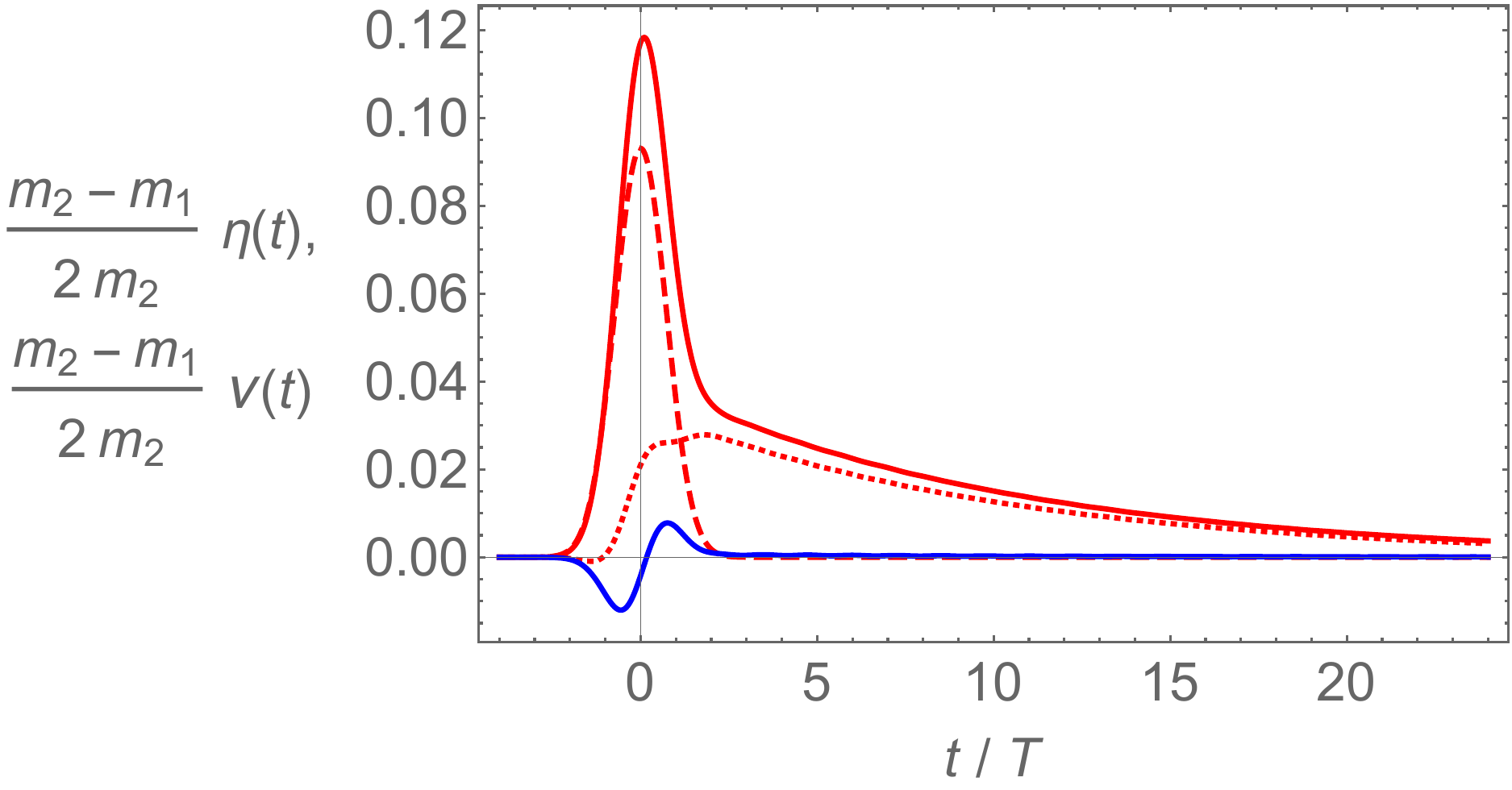} 
   \caption{The  relative variations of the effective plasma frequency and the loss rate, $\eta$ and $\nu$, scaled by
  $\left(1 - m_1^*/m_2^*\right)/2$, induced by the Gaussian optical pump pulse $I_p\left(t\right) = I_0 \exp\left( - t^2 / T^2\right)$,
  for $\alpha_* I_0 =0.1$, $\tau_1 = 10 T$, $\tau_2 = 5T$, and  $\omega_S T = 5$. Solid red line shows $\eta\left(t\right)$, 
  while dashed and dotted red curves correspond respectively to $\eta_1$ and $\eta_2$. Blue line represents $\nu\left(t\right)$.
   }
   \label{fig:epsilon}
 \end{figure}

In Fig. \ref{fig:epsilon}, we show the time dependence of the relative changes in the effective plasma frequency and in the 
effective loss rate. Note the expected two-stage evolution in both of these parameters -- a rapid change for the 
duration of the pump pulse due to the virtual interband transitions, followed by effect of multi-photon absorption that persist on the time scale of the 
interband relaxation time $\tau_1$. 

\section{Ultrafast virtual modulation in transparent conducting oxides}

With the plasma frequency in the near-infrared, and absorption edge at the higher frequency end or even beyond the visible range,\cite{TCOs}
transparent conducting oxides offer a convenient platform for the implementation of the proposed approach to ultrafast optical modulation by 
virtual interband transitions. In particular, for the indium-tin oxide (ITO) the absorption edge $\hbar \omega_c$ 
corresponds  to the blue portion of the visible spectrum, while the plasma frequency is in the infrared range.\cite{ITO} 
The necessary requirements for the implementation of  our approach (\ref{eq:regime}) can therefore be satisfied  with the use of a few-cycle pump pulse at lower visible frequencies and the signal in the near-infrared range.

As the bandstructure of the ITO \cite{Julia} corresponds to the case (b) of Fig. \ref{fig:bands}, 
with the effective masses $m_1^* > 0$ and $m_2^* < 0$ Eqns. (\ref{eqn:epsilon})-(\ref{eqn:eta2})  imply that  the
the dielectric permittivity of the ITO will increase with the arrival of the optical pump pulse, and decrease at its departure. 
If another optical signal is simultaneously
propagating in the modulated material (see Fig. \ref{fig:schematics}), it will then experience red frequency shift when the 
permittivity goes up, and blue shift when it
goes down. 

This physical mechanism explains the origin of such behavior recently found in the experiments of Ref. \cite{VSMoti}, where ITO thin film was simultaneously illuminated by 
high-intensity femtosecond pulses at the (central) wavelength of 800 nm, and slowly varying ``probe'' beam at $1200$ nm wavelength -- leading to the observation of 
red frequency shift immediately followed by the blue frequency shift. As the experimental parameters in Ref.  \cite{VSMoti} are well  within the constraints of the inequality  (\ref{eq:regime}),
that work should be considered as the world's first experimental implementation of the proposed ultrafast optical modulation by virtual interband transitions.

We however emphasize, that our approach is not in any way special to ITO or even to the whole group of transparent conducting oxides, but with the proper choice
of experimentally controlled parameters (optical pump pulse duration $T$ and central  frequency $\omega_P$, and the signal frequency $\omega_S$) it can be applied to any optical material supporting free carriers.

\section{Conclusions}

We have introduced a new approach to ultrafast optical modulation, based on the virtual interband transition excitation. Being inherently ultrafast and relaxation-free, it offers a
direct and practical way to create ``temporal boundaries'' for light, in already existing materials and with a straightforward implementation  using only the standard toolbox of modern ultra-fast optics -- thus finally opening the route to the elusive Photonic Time Crystal at optical frequencies.

\section{Acknowledgements}

The author would like to thank Profs. Sasha  Boltasseva and Vlad Shalaev  for helpful discussions. 



\begin{thebibliography}{10}

\bibitem{BornWolf}
M. Born and E. Wolf, Principles of Optics: Electromagnetic Theory of Propagation, Interference and Diffraction of Light, 7th ed. (Cambridge, Cambridge University Press, 1999).

\bibitem{Boyd}
R. Boyd, {\it Nonlinear Optics} (Academic Press; 2nd edition, 2002).

\bibitem{Khurgin-index} 
J. B. Khurgin, ``Energy and Power requirements for alteration of the refractive index,''
preprint arXiv:2308.16011 (2023).

\bibitem{strain}
C. E. Campanella, A. Cuccovillo, C. Campanella, A. Yurt, and V. M. Passaro, ``Fibre Bragg grating
based strain sensors: Review of technology and applications,"  Sensors {\bf 18}, 3115 (2018).
 
\bibitem{acousto-optics} 
A. Korpel, ``Acousto-optics?a review of fundamentals," Proceedings of the IEEE {\bf 69}, 48 - 53 (1981).

\bibitem{thermal} 
J. Komma, C. Schwarz, G. Hofmann, D. Heinert, and R. Nawrodt, "Thermo-optic coefficient of silicon at 1550 nm and cryogenic temperatures," Applied Physics Letters {\bf 101}, 041905 (2012).

\bibitem{electrooptics} 
R. Soref and B. Bennett, ``Electrooptical effects in silicon," IEEE Journal of Quantum Electronics {\bf 23},
123 - 129 (1987).

\bibitem{PTC1}
 J. R. Zurita-S\'anchez, J. H. Abundis-Pati\~no, and P. Halevi, ``Pulse propagation through a slab with time-periodic dielectric function $\epsilon(t)$,'' Opt. Express {\bf 20},  5586 - 5600 (2012).

\bibitem{PTC2}
E. Lustig, Y. Sharabi, and M. Segev,``Topological aspects of photonic time crystals,'' Optica {\bf 5}, 1390 (2018).

\bibitem{VSMotiR1}
E. Lustig, O. Segal, S. Saha, C. Fruhling, V. M. Shalaev, A. Boltasseva, and M. Segev, ``Photonic time-crystals - fundamental concepts,'' 
Opt. Express {\bf 31}, 9165 - 9170 (2023). 

\bibitem{VSMotiR2}
S. Saha, O. Segal, C. Fruhling, E. Lustig, M. Segev, A. Boltasseva, and V. M. Shalaev, ``Photonic time crystals: a materials perspective,'' 
Opt. Express {\bf 31}, 8267 - 8273 (2023).

\bibitem{LLQM}
 L. D. Landau and  L. M. Lifshitz,  {\it Quantum Mechanics}, (Butterworth-Heinemann; 3rd edition, 1981).

\bibitem{LLECM} 
L. D. Landau, L. P. Pitaevskii, and E. M. Lifshitz, {\it Electrodynamics of Continuous Media} 
(Butterworth-Heinemann; 2nd edition, 1984).

\bibitem{SCai}
W. Cai and  V. Shalaev, \textit{Optical metamaterials: fundamentals and applications} (Springer Science \& Business Media, 2009).

\bibitem{TCOs}
W. Jaffray, S. Saha, V. M. Shalaev, A. Boltasseva, and M. Ferrera, ``Transparent conducting oxides: from all-dielectric plasmonics to a new paradigm in integrated photonics,''  Advances in Optics and Photonics {\bf 14} (2),  148 - 208 (2022).

\bibitem{nmat}
A.~J.~Hoffman, L. Alekseyev, S. S. Howard, K. J. Franz, D. M. Wasserman, V. A. Podolskiy, E. E. Narimanov, D. L. Sivco, C. Gmachl, "Negative refraction in semiconductor metamaterials," Nature Materials {\bf 6}, 948 (2007).

\bibitem{Yamanishi-PRL-1987}
M. Yamanishi, ``Field-Induced Nonlinearity Due to Virtual Transitions in Semiconductor Quantum-Well Structures,''
Phys. Rev. Lett. {\bf 59} (9), 1014 - 1017  (1987).

\bibitem{SRC-PRL-1986}
S. Schmitt-Rink and D. S. Chemla, 
``Collective Excitations and the Dynamical Stark Effect in a Coherently Driven Exciton System,''
Phys. Rev Lett. {\bf 57}, 2752 (1986).

\bibitem{CMSR-PRL-1987}
D. S. Chemla, D. A. B. Miller, and  S. Schmitt-Rink, 
``Generation of Ultrashort Electrical Pulses through Screening by Virtual Populations in Biased Quantum Wells,"
Phys. Rev Lett. {\bf 59}, 1018 (1987).

\bibitem{Mysyrowicz-PRL-1986}
A. Mysyrowicz, D. Hulin, A. Antonetti, A. Migus, W. T. Masselink, and H. Morkoc,
``Dressed Excitons" in a Multiple-Quantum-Well Structure: Evidence for an Optical Stark Effect with Femtosecond Response Time,''
Phys. Rev. Lett. {\bf 56}, 2748 (1986).

\bibitem{VLCZH-OL-1986}
A. Von Lehmen, D. S. Chemla, J. E. Zucker, and J. P. Heritage
``Optical Stark effect on excitons in GaAs quantum wells,''
Opt. Lett. {\bf 11} (10), 609 - 611 (1986).  

\bibitem{Ziman}
 J. M. Ziman, ``Electrons and Phonons: The Theory of Transport Phenomena in Solids,''  Oxford University Press, Reprint edition, 2001.

\bibitem{percolation}
A. Sihvola, S. Saastamoinen and K. Heiska, Remote Sensing Reviews {\bf 9}, 35 (1994).

\bibitem{Julia}
J.E. Medvedeva,
``Magnetically Mediated Transparent Conductors: In$_2$O$_3$ doped with Molybdenum,''
Phys. Rev. Lett. {\bf 97}, 086401 (2006). 
 
 \bibitem{ITO}
 T. S. Kim, C. H. Choi, T. S. Jeong and K. H. Shim,
``Influence of Substrate Temperature and O$_2$ Flow on the Properties of
RF-Magnetron-Sputtered Indium-Tin-Oxide Thin Films,''
 Journal of the Korean Physical Society {\bf  51} (2),  534 - 538 (2007).

 \bibitem{VSMoti} 
E. Lustig, O. Segal, S. Saha, E. Bordo, S. N. Chowdhury, Y. Sharabi , A. Fleischer , A. Boltasseva, O. Cohen , V. M. Shalaev and M. Segev,
``Time-refraction optics with single cycle modulation,''   Nanophotonics {\bf  12} (12), 2221 - 2230 (2023).
  
\end{thebibliography}
\end{document}